\documentclass[11pt]{article}
\usepackage[english]{babel}
\usepackage{amsmath,amssymb,graphicx,xcolor}
\usepackage{fullpage}
\usepackage{authblk}

\title{Photo-Motile Structures}
\author[1]{Kevin Korner}
\author[2]{Basile Audoly}
\author[1]{Kaushik Bhattacharya}

\affil[1]{Division of Engineering and Applied Science, California Institute of Technology,  USA}
\affil[2]{Laboratoire de m{\'e}canique des solides, CNRS, Institut Polytechnique de Paris, Palaiseau, France}

\date{\today}

\graphicspath{{figures/}}

\begin{document}

\maketitle

\textbf{{\small {\noindent}Actuation remains a significant challenge in soft
robotics. Actuation by light has important advantages: objects can be
actuated from a distance, distinct frequencies can be used to actuate and
control distinct modes with minimal interference and significant power can be
transmitted over long distances through corrosion-free, lightweight fiber
optic cables. Photochemical processes that directly convert photons to
configurational changes are particularly attractive for actuation. Various
researchers have demonstrated light-induced actuation with liquid crystal
elastomers combined with azobenzene photochromes. We present a simple modeling
framework and a series of examples that studies actuation by light. Of
particular interest is the generation of cyclic or periodic motion under
steady illumination. We show that this emerges as a result of a coupling
between light absorption and deformation. As the structure absorbs light and
deforms, the conditions of illumination change, and this in turn changes the
nature of further deformation. This coupling can be exploited in either closed
structures or with structural instabilities to generate cyclic motion. }}


\paragraph{Significance} {\it Actuation and propulsion are significant
challenges in soft robotics.  Supply of power typically requires a
cumbersome tether or heavy on-board power source.  Further, one
typically needs to reset the system.  In this work, we show that this
challenge can be overcome by the use of photo-mechanical materials and
actuation by light.  We develop a simple modeling framework which
reveals how steady illumination from a distance can give rise to
cyclic motion.  Such motion can be exploited for actuation and
propulsion with no need for tether or on-board power source, through
the natural but nonlinear/non-local coupling between deformation and
light absorption.}

\newpage A major challenge in soft robotics is the integration of
sensing, actuation, control, and propulsion.  In most soft robotic
systems, propulsion and controls are enabled through a physical tether
or complex on-board electronics and batteries.  A tether simplifies
the design but limits the range of motion of the robot, while on-board
controls and power supplies can be heavy and can complicate the design
\cite{cbsd_pnas_18}.  Actuation by
light through photomechanical processes directly convert photons to
deformation and offers an attractive alternative.  It can deliver
energy remotely.  Further, multiple frequencies can be used to actuate
and sense different modes.  Finally, if a tether is an option, then a
significant energy can be delivered through corrosion-free and
lightweight fiber-optic cables.

A further challenge arises in propulsion where one needs to generate cyclic
motion. Since most actuation systems actuate one way, there is a need to reset
the system~{\cite{cbsd_pnas_18}}. To simplify the control process, it is
desirable to do so by inherent response rather than by pulsing of the external
source. Actuation by light is again attractive because one can use the
directionality of the propagation of light. As the structure absorbs light and
deforms, the conditions of illumination change, and this in turn changes the
nature of further deformation. This coupling can be exploited in either closed
structures or with structural instabilities to generate cyclic motion.

These advantages have motivated a recent body of work on developing
photomechanical materials (see~{\cite{w_book_17}} for an extensive review).
Much of this work has focussed on incorporating azobenzene photochromes that
absorb light and transform between {\emph{cis}} and {\emph{trans}}
configurations into liquid crystal elastomers whose orientational order is
coupled to deformation, following the pioneering work of Yu \emph{et
al.}~{\cite{yni_nat_03}}. These materials are typically synthesized as thin
strips which bend when illuminated with light of appropriate frequency.
Further they can be combined with structural polymers to provide robustness
{\cite{Gelebart2017}}.

Various works have demonstrated the ability to generate cyclic motion under
steady illumination. Yamada \emph{et al.}~{\cite{Yamada2008}} demonstrated
that a ring of LCE film containing azobenzene derivatives can roll in the
presence of illumination. When wrapped around a series of pulleys, the film
can be used as a light-driven plastic motor system. White \emph{et
al.}~{\cite{wetal_sm_08}} developed a high frequency oscillator from a strip
which bends under illumination sufficiently to block the light source and
reset. Wei \emph{et al.}~{\cite{Wie2016}} produced rolling motion in
monolithic polymer films where ultraviolet-visible light transforms the film
from flat sheets to spiral ribbons, which then rolls under continuous
illumination. Finally, Gelebart \emph{et al.}~{\cite{Gelebart2017}}
created an oscillatory behavior of a doubly clamped LCE film.

In this paper, we develop a theoretical framework to understand
actuation deformation by light, and study a series of examples
motivated by the experiments described above.  Modeling light-mediated
actuation is a complex multiphysics process involving three key
elements: propagation and absorption of light, chemical transformation
and temporal evolution of chromophores between states, and the
large-deformation nonlinear mechanics of deformation.  Corbett and
Warner developed a theory of light absorption and actuation in
azobenzene containing liquid crystal elastomers~{\cite{cw_prl_06}},
used it to study the large deformation of illuminated thin strips
(planar elastica model)~{\cite{Corbett2015}}.  We build on that work
with a focus on cyclic or periodic motion under steady illumination.
Our resulting model is quite simple and can be solved numerically in
real time on any personal computer, while capturing a rich range of
behaviors.  It is highly effective in revealing the underlying
mechanisms, as we demonstrate through examples, and can be used as a
tool for the design and control of this novel type of structures.

\section{Photo-deformable elastica}\label{sec: PhotoElasticity}

\begin{figure}
  \begin{center}
    \includegraphics[width=3in]{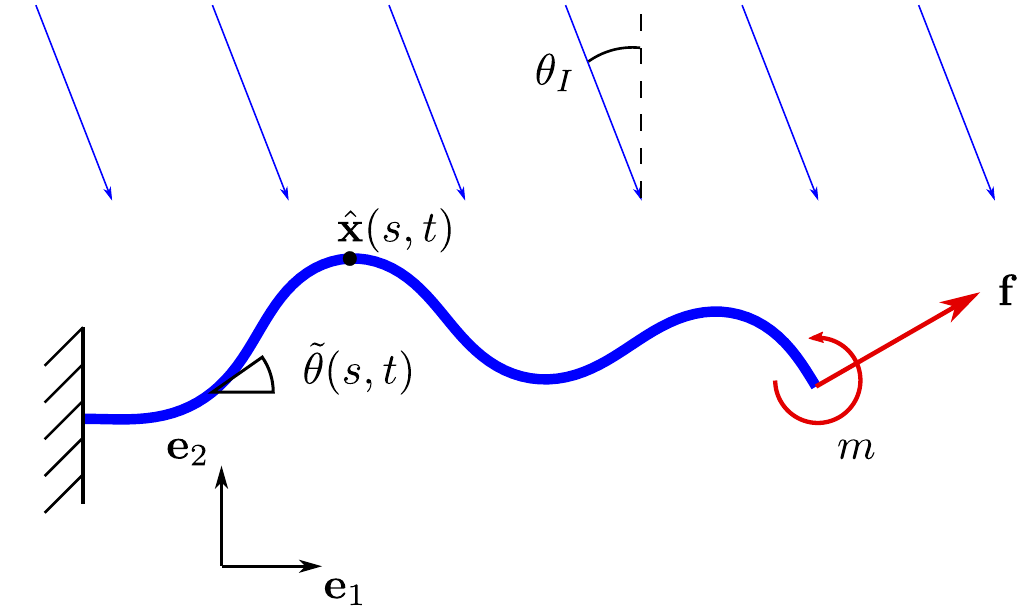}
  \end{center}
  \caption{Elastica under illumination.\label{fig:elastica}}
\end{figure}

Consider an inextensible beam or a strip (planar elastica) subjected
to illumination as shown in Figure~\ref{fig:elastica}.  Let
$\mathbf{x} (s, t)$ denote the position of centerline point $s$ at
time $t$ and $\theta (s, t)$ denote the angle that the tangent to the
beam makes with the horizontal axis $\mathbf{e}_1$.  We assume that
the deformation caused by illumination takes place over a
significantly slower time scale than the natural periods of the beam
so that we may assume that the beam is at equilibrium at all times.
Therefore, at each $t$,
\begin{eqnarray}
  \frac{\partial \mathbf{f}}{\partial s} (s, t) & = & 0,  \label{eq:eqf}\\
  \frac{\partial m}{\partial s} (s, t) + (\widehat{\mathbf{t}} (s, t) \times
  \mathbf{f} (s, t)) \cdot \mathbf{e}_3 & = & 0  \label{eq:eqm}
\end{eqnarray}
where $\widehat{\mathbf{t}} (\theta (s, t)) = \partial \mathbf{x} /
\partial s (s, t) = \cos \theta (s, t) \mathbf{e}_1 + \sin \theta (s,
t) \mathbf{e}_2$ is the unit tangent, $\mathbf{f} (s, t)$ is the
internal force transmitted across a cross-section, and $m (s, t)$ is
the internal moment about $\mathbf{e}_3$.

Since we assume that the beam is inextensible and unshearable, the internal
force $\mathbf{f}$ is constitutively indeterminate and we only need to specify
a constitutive law for the moment $m$. Following Corbett and Warner
{\cite{Corbett2015}}, we assume that the beam is made of an elastic material
whose spontaneous or stress-free strain, $\varepsilon_0$, changes with time
depending on local population of \emph{cis} molecules. The longitudinal
stress at a point at a position $s$ along the length of the beam, $z$ along
the depth of the beam and at time $t$ is given by Hooke's law, $\sigma (s, z,
t) = E (\varepsilon (s, z, t) - \varepsilon_0 (s, z, t))$, where $\varepsilon$
is the strain and $\varepsilon_0$ is the spontaneous strain. The moment is
found by integration through the thickness as
\begin{equation}
  \label{eq:m} m (s, t) = \int_{- h / 2}^{h / 2} E (\varepsilon (s, z, t) -
  \varepsilon_0 (s, z, t)) z \mathrm{d} z
\end{equation}
with $z = 0$ taken to be the center of the beam.  The strain is
related to curvature as in classical elastica theory\footnote{We
assume that the neutral axis is unaffected by illumination since the
penetration depth is small, as argued later.}, and the spontaneous
strain depends on the built-in curvature $\kappa_r$ of the beam (the
curvature with no applied load and no illumination) and the
concentration $n_c$ of the \emph{cis} molecules:
\begin{eqnarray}
  \varepsilon (s, z, t) & = & \kappa (s, t) z,  \label{eq:e}\\
  \varepsilon_0 (s, z, t) & = & \kappa_r (s) z - \lambda n_c (s, z, t) 
  \label{eq:e0}
\end{eqnarray}
where $\lambda$ is a constant of proportionality linking the longitudinal
strain and concentration of \emph{cis} molecules.   $\lambda >
0$  when the \emph{cis} molecules induce an expansion, while
$\lambda < 0$ corresponds to an induced contraction.

Substituting~(\ref{eq:e}) and~(\ref{eq:e0}) into~(\ref{eq:m}), we find the
constitutive law in the form
\begin{equation}
  \label{eq:mconst} m (s, t) = \frac{Eh^3}{12}  (\kappa (s, t) - \kappa_0 (s,
  t))
\end{equation}
where
\begin{equation}
  \label{eq:k0} \kappa_0 (s, t) = \kappa_r (s) - \frac{12 \lambda}{h^3} 
  \int_{- h / 2}^{h / 2} n_c (s, z, t) z \mathrm{d} z.
\end{equation}
It remains to specify the evolution of the spontaneous curvature in
the presence of illumination.  The concentration of \emph{cis}
molecules is increased by photon absorption, and decreased by thermal
decay:
\[ \tau \dfrac{\partial n_c}{\partial t} (s, z, t) = - n_c (s, z, t) + (1 -
   n_c (s, z, t)) \alpha_1 \mathcal{I} (s, z, t), \]
where $\alpha_1$ is a material constant and $\mathcal{I} (s, z, t)$ denotes
the illumination, {\emph{i.e.}}, the quantity photons per unit time arriving
at the depth $z$ at time $t$. In typical materials, $n_c \ll
1$ is small \cite{w_book_17} and we can
simplify the differential equation to
\begin{equation}
  \label{eq:dncdt} \tau \dfrac{\partial n_c}{\partial t} (s, z, t) = - n_c
  (s, z, t) + \alpha_1 \mathcal{I} (s, z, t) .
\end{equation}
Further, we assume that the illumination changes slowly since deformation is
slow, and that the optical penetration depth $d$ is small since the absorption
spectra of the both the \emph{cis} and \emph{trans} states typically
overlap and the proportion of \emph{cis} molecules is small \cite{Yu2004,w_book_17}. Then, the
illumination follows Beer's law,
\begin{equation}
  \label{eq:ill} \mathcal{I} (s, z, t) =\mathcal{I} \left( s, \frac{h}{2}, t
  \right) \exp \left( - \frac{h / 2 - z}{d} \right)
\end{equation}
where $z = h / 2$ is the free surface that is illuminated\footnote{Note that the 
result (\ref{eq:k0evol}) does not require the exponential profile of Beer's 
law, but simply a steady profile, $\mathcal{I} (s, z, t) =\mathcal{I}_0 \left( s, t \right) f(z)$.}
. Combining (\ref{eq:k0}),
(\ref{eq:dncdt}) and~(\ref{eq:ill}),
\[ \tau \frac{\partial \kappa_0}{\partial t} (s, t) = \frac{12 \lambda}{h^3} 
   \int_{- h / 2}^{h / 2} \tau \frac{\partial n_c}{\partial t} (s, z, t) z
   \mathrm{d} z = - (\kappa_0 (s, t) - \kappa_r (s)) + \alpha \mathcal{I} \left(
   s, \frac{h}{2}, t \right) \]
where $\alpha = -\frac{12 \lambda \alpha_1}{h^3} \int_{- h / 2}^{h /
2} \exp \left( - \frac{h / 2 - z}{d} \right) z \mathrm{d} z$ is an
effective (macroscopic) coupling constant.  Finally, the absorption of
light on the surface depends on light intensity $I_0$ and on the
relative orientation of the light and the strip, $\mathcal{I} \left(
s, \frac{h}{2}, t \right) = I_0 f (\theta (s, t) - \theta_I)$, where
$\theta_I$ is the angle of illumination.  Therefore,
\begin{equation}
  \label{eq:k0evol} \tau \frac{\partial \kappa_0}{\partial t} (s, t) +
  (\kappa_0 (s, t) - \kappa_r (s)) = \alpha I_0 f (\theta (s, t) - \theta_I) .
\end{equation}
If we ignore shadowing, then it is reasonable to assume that the $f$ is the
apparent area normal to light propagation,
\begin{equation}
  f (\phi) = \left\{\begin{array}{ll}
    \cos (\phi) & \text{if } \phi \in (- \pi / 2, \pi / 2),\\
    0 & \text{else} .
  \end{array}\right.
\end{equation}
Finally, we combine~(\ref{eq:eqf}), (\ref{eq:eqm}), (\ref{eq:e})
and~(\ref{eq:mconst}), and non-dimensionalize the resulting equation along
with~(\ref{eq:k0evol}), introducing the scaled arclength $S = s / l$ (where
$l$ is the length of the beam), the scaled time $T = t / \tau$ and the scaled
curvature $K = l \kappa$,
\begin{eqnarray}
  \frac{\partial}{\partial S}  \left( \frac{\partial \theta}{\partial S} (S,
  T) - K_0 (S, T) \right) - F_x \cos \theta (S, T) + F_y \sin \theta (S, T) = 0,  \label{eq:equil}\\
  \frac{\partial K_0}{\partial T} (S, T) + (K_0 (S, T) - K_r (S)) =
  \Lambda f (\theta (S, T) - \theta_I) .  \label{eq:evol}
\end{eqnarray}
Here, the scaled light intensity $\Lambda = \alpha lI_0$ is the only
non-dimensional parameter of the problem and the components of the internal
force $F_x, F_y$ are yet unknown constants. Based on the estimates given in
Table~\ref{tab:par}, the parameter $|\Lambda| \sim 2.4$ is of order 
one. To predict how the shape of the beam evolves with time, we
solve these equations~(\ref{eq:equil}) and~(\ref{eq:evol}) for $\theta (S, T)$
using a numerical method described in the appendix with specific initial,
boundary and illumination conditions.

For future reference, we note that the equilibrium
equation~(\ref{eq:equil}) can be derived by the Euler-Lagrange method
as the stationarity condition of the energy functional
\begin{equation} \label{eq:be}
  \mathcal{E} [\theta] = \int_0^1 \frac{1}{2}  \left| \frac{\partial
  \theta}{\partial S} - K_0 \right|^2 \mathrm{d} S.
\end{equation}

\begin{table}
  \begin{center}
    \begin{tabular}{|c|c|}
      \hline
      Parameter & Typical Value\\
      \hline
      $\alpha_1$ & $0.001$ {\cite{pc}}\\
      $I_0$ & $100 \text{W/m}^2$ {\cite{Smith2014}}\\
      $\lambda$ & $-1 / 20$ {\cite{Corbett2015}}\\
      $E$ & $0.6 - 4$ GPa {\cite{Smith2014}}\\
      $h$ & $15 \mu$\text{m} {\cite{Smith2014}}\\
      $d$ & $1.5 \mu$\text{m} {\cite{pc}}\\
      $w$ & $1$ mm {\cite{Smith2014}}\\
      $l$ & $15$ mm {\cite{Smith2014}}\\
      \hline
    \end{tabular}
  \end{center}
  \caption{Estimates of the experimental parameters based on the
  literature.\label{tab:par}}
\end{table}

\section{Rolling ring}\label{sec:rolling}

\begin{figure}
  \begin{center}
    \includegraphics[width=6.5in]{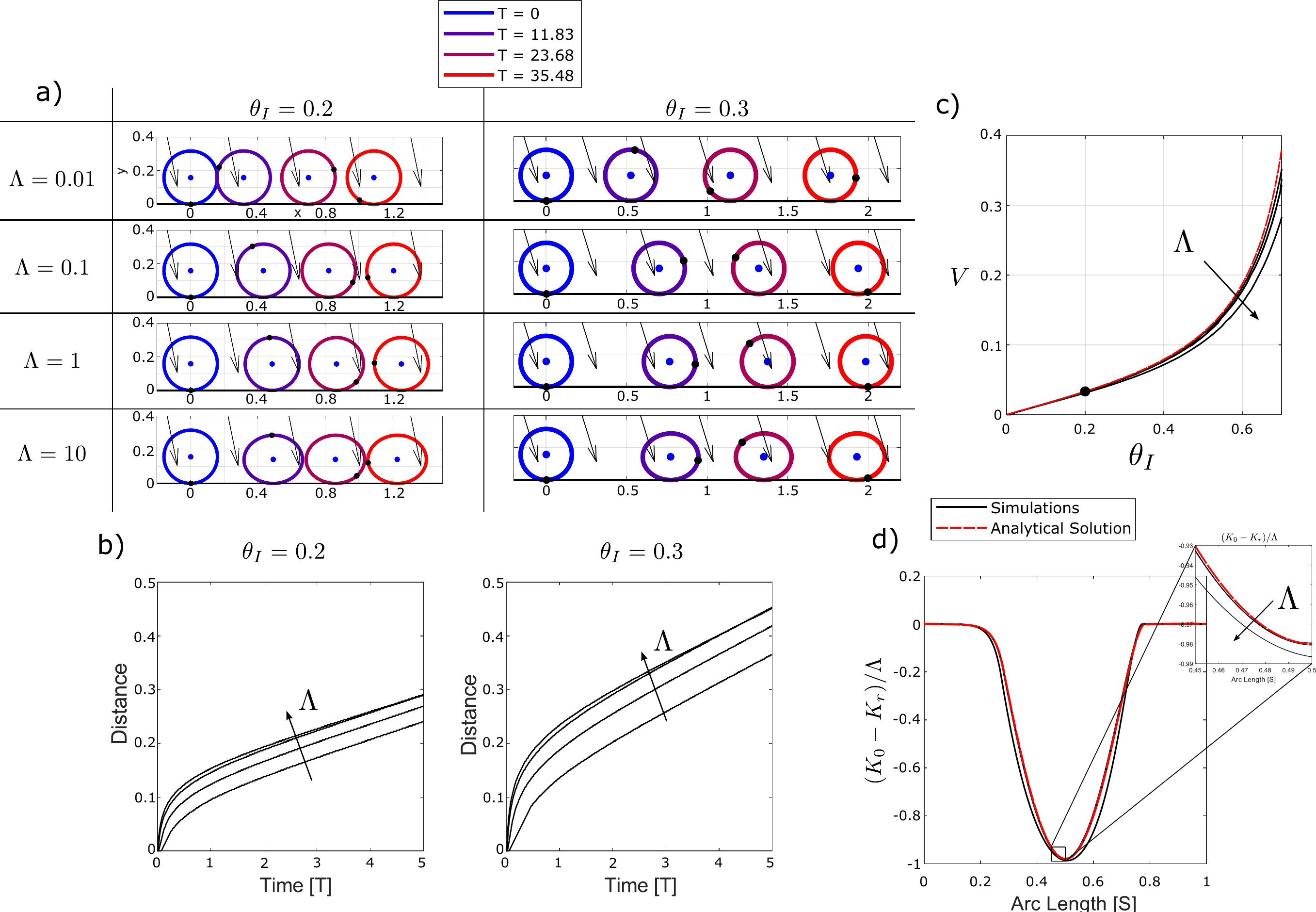}
  \end{center}
  \caption{Rolling ring.  (a)~Snapshots of an initially
  circular ring with radius $R=1/(2\pi)$ subjected
  to illumination at angle $\theta_I$ and of intensity $\Lambda$ at
  times $T = \{0, 11.83, 23.68, 35.48\}$.  The point that is initially
  in contact with the ground is marked with a black dot while the center of mass is the blue dot.
  (b)~Distance travelled by the rolling ring vs.~time for various
  intensities $\Lambda = \{0.01, 0.1, 1, 10\}$.  Note that the a
  steady velocity is reached in all cases, after an initial transient.
  (c)~Steady state velocity as a function of illumination angle and
  intensity.  The velocity increases when the illumination angle
  moves away from the vertical, but is relatively insensitive to the
  intensity of illumination.  (d)~Scaled change of spontaneous
  curvature induced by illumination along the beam for $\theta_I = 0.2$ (indicated by dot in (c)), for various
  illumination intensities.  This quantity appears to be largely
  insensitive to the intensity of illumination.  Simulation data is
  shown as solid black lines while the analytical solution given by
  solving Equation (\ref{eq: H}) is shown as a red dashed line. \label{fig:rolling}}
\end{figure}

Our first example is motivated by the work of Yamada \emph{et
al.}~{\cite{Yamada2008}} on a rolling ring and motor, as well as that of Wei 
\emph{et al.}~{\cite{Wie2016}} on a rolling spiral. We consider a closed, initially
circular ring on a rigid horizontal surface, which is illuminated with a
steady source at angle $\theta_I$. The fact that the ring is closed implies
that
\begin{equation}
  \label{eq:closed} \int_0^1 \sin \theta (S, T) \mathrm{d} S = \int_0^1 \cos
  \theta (S, T) \mathrm{d} S = 0,
\end{equation}
as well as $\theta (0, T) = \theta (1, T)$. We assume that the ring makes a
tangential rolling contact with the horizontal surface so that $X (S_c (T), T)
= S_c (T)$, $Y (S_c (T), T) = 0$ and
\begin{equation}
  \label{eq:contact} \theta (S_c (T), T) = 0,
\end{equation}
where $S_c (T)$ is the point of contact. We determine this point of contact by
assuming overall mechanical equilibrium of the ring under gravity so that the
center of mass of the ring is always vertically above the point of contact,
\begin{equation}  \label{eq: rollingconditiongeneral}
  \begin{split}
    S_c (T) &= X (S_c (T), T) = \int_0^1 X (S, T) \mathrm{d} S = \int_0^1 \left(
    \int_0^S \cos \theta (\tilde{S}, T) \mathrm{d} \tilde{S} \right) \mathrm{d} S\\
    &= \int_0^1 (1 - S) \cos \theta (S, T) \mathrm{d} S = -
    \int_0^1 S \cos \theta (S, T) \mathrm{d} S.
  \end{split}
\end{equation}
We set $K_r = 2 \pi$ and $\theta (S, 0) = 2 \pi S$ corresponding to an
initially circular ring and solve the equations~(\ref{eq:equil}),
(\ref{eq:evol}) subject to the conditions above.
Figure~\ref{fig:rolling}(a) shows snapshots of the ring for various
angles and intensity of illumination.  In each case, the ring deforms
as it is illuminated, in a way which is non-symmetric with respect to
the vertical axis and depends on the angle of illumination.  This
asymmetry causes the center of mass of the ring to move, which in turn
causes the ring to roll.  Figure~\ref{fig:rolling}(b) shows the
distance travelled by the point of contact as a function of time under
various angles and intensity of illumination.  After an initial
transient, the ring rolls with a steady velocity and has an invariant
shape.  The steady velocity is plotted as a function of the
illumination angle for various illumination intensities in
Figure~\ref{fig:rolling}(c): it is zero when the illumination is
vertical ($\theta_I = 0$), which is a consequence of the symmetry, and
increases with increasing angle of illumination $\theta_I$.
Remarkably, \emph{the rolling velocity is practically independent of
the intensity of illumination} in the range of values of $\Lambda$
relevant to the experiments and investigated here.  To investigate
this further, we plot the scaled deviation in spontaneous curvature
$(K_0 - K_r) / \Lambda$ as a function of arclength in
Figure~\ref{fig:rolling}(d): this quantity appears to be practically
independent of the intensity of illumination as well.  This shows that
amount of deformation scales linearly with the light intensity, while
the deformation mode (and, hence, the asymmetry and the rolling
velocity) is largely independent of the intensity.

To understand these features, we analyze steadily rolling solutions,
{\emph{i.e.}}, we seek solutions of the form $\theta (S, T) = \Theta
(S - VT)$ and aim at identifying the rolling velocity $V$.  We set
$\omega = 2 \pi (S - VT)$ choosing $T = 0$ to be a time when the point
in contact with the ground is $S = S_c (0) = 0$.  This implies
\begin{equation}
  \label{eq:thetazero} \Theta (0) = 0.
\end{equation}
The rolling condition~(\ref{eq: rollingconditiongeneral}) becomes
\begin{equation}
  \label{eq:rolling} 0 = \int_0^{2 \pi} \omega \cos \Theta (\omega) \mathrm{d}
  \omega,
\end{equation}
and the evolution equation~(\ref{eq:evol})
\begin{equation}
  \label{eq:progressiveK0} - 2 \pi V \dfrac{\mathrm{d} K_0}{\mathrm{d} \omega} + (K_0
  - 2 \pi) = \Lambda f (\Theta - \theta_I) .
\end{equation}
We now assume that \emph{the shape of the ring is almost circular} so that
\begin{equation}
  \label{eq:lin} \Theta (\omega) = \omega + \Theta_1 (\omega), \quad K_0
  (\omega) = 2 \pi + K_1 (\omega)
\end{equation}
where $| \Theta_1 | \ll 1$ and $|K_1 | \ll 1$ are treated as perturbations.
Keeping only terms linear in $\Theta_1, K_1$, the equilibrium
equation~(\ref{eq:equil}) and closure condition~(\ref{eq:closed}) become
\begin{equation}
  \begin{array}{c}
    4 \pi^2 \Theta_1'' (\omega) - 2 \pi K_1' (\omega) + F_y \cos \omega - F_x
    \sin \omega = 0 \hspace{0.17em},\\
    \int_0^{2 \pi} \cos (\omega) \Theta_1 (\omega) \mathrm{d} \omega = \int_0^{2
    \pi} \sin (\omega) \Theta_1 (\omega) \mathrm{d} \omega = 0 \hspace{0.17em} .
  \end{array} \label{eq: linearizationproblem}
\end{equation}
Introducing the Fourier transform $\hat{f} (k) = \int_0^{2 \pi} f (\omega)
\exp (- ik \omega) \mathrm{d} \omega$ where $k$ is an integer, we can
solve~(\ref{eq:progressiveK0}) as
\begin{equation}
  \label{eq: K1def} \hat{K}_1 (k) = \frac{\Lambda \hat{f}_I (k)}{1 - 2 i \pi
  kV} \hspace{0.17em},
\end{equation}
where
\begin{equation}
  f_I (\omega) = f (\omega - \theta_I) . \label{eq:fI}
\end{equation}
Similarly, we can solve equation~(\ref{eq: linearizationproblem}) in Fourier
form as
\begin{equation}
  \begin{array}{lllll}
    \hat{\Theta}_1 (\pm 1) & = & 0 & {} {} \qquad & \text{for
    $| k | = 1$},\\
    \hat{\Theta}_1 (k) & = & - i \frac{\hat{K}_1 (k)}{2 \pi k} &  & \text{for
    $| k | \geqslant 2$}.
  \end{array} \label{eq:ThetaHatButZero}
\end{equation}
Note that the first equation in~(\ref{eq: linearizationproblem}) yields $F_x$
and $F_y$ in terms of $\hat{\Theta}_1 (\pm 1)$ and $\hat{K}_1 (\pm 1)$ as
well, but these expressions are not needed.

The horizontal tangency condition~(\ref{eq:thetazero}) reads $0 = \Theta (0) =
\Theta_1 (0) 
= \frac{1}{2 \pi}  \sum_k \hat{\Theta}_1 (k)$ where the sum runs over all
{\emph{signed}} integers $k$. Rearranging the terms in the sum and solving for
$\hat{\Theta}_1 (0)$, we find
\begin{equation}
  \hat{\Theta}_1 (0) = - 2 \sum_{k \geqslant 1} \operatorname{Re} \hat{\Theta}_1 (k)
  \label{eq:thetaHat1Zero}
\end{equation}
where we have used $\hat{\Theta}_1 (- k) + \hat{\Theta}_1 (k) =
\overline{\hat{\Theta}_1 (k)} + \hat{\Theta}_1 (k) = 2
\operatorname{Re} \hat{\Theta}_1 (k)$ since $\Theta_1 (\omega)$ is a
real function.  Here, $\overline{z}$ denoting the conjugate of the
complex number $z$.

Equations~(\ref{eq: K1def}--\ref{eq:thetaHat1Zero}) yield the shape in terms
of the known illumination parameter $\Lambda$ and of the unknown scaled
rolling velocity $V$. The latter can be found by linearizing the rolling
condition~(\ref{eq:rolling}) as $\int_0^{2 \pi} g (\omega) \Theta_1 (\omega) d
\omega = 0$ where $g (\omega) = \omega \sin \omega .$ Using Parseval's
identity, this can be rewritten as
\begin{equation}
  \frac{1}{2 \pi}  \sum_k \hat{g} (k)  \hat{\Theta}_1 (- k) = 0, \qquad
  \text{where }  \hat{g} (k) = \left\{\begin{array}{lll}
    - \frac{\pi}{2}  (2 \pi ik + 1) & {} {} \quad & \text{if }
    |k| = 1 \hspace{0.17em},\\
    \frac{2 \pi}{k^2 - 1} &  & \text{if } |k| \neq 1 \hspace{0.17em} .
  \end{array}\right.
\end{equation}
Inserting~(\ref{eq:ThetaHatButZero}--\ref{eq:thetaHat1Zero}) into this
equation, we obtain $2 \sum_{k \geqslant 2} \frac{k^2}{k^2 - 1} \operatorname{Re}
\hat{\Theta}_1 (k) = 0$ which, in view of~(\ref{eq:
K1def}--\ref{eq:ThetaHatButZero}), yields an implicit equation for the rolling
velocity $V$ in terms of the angle of illumination $\theta_I$,
\begin{equation}
   \Lambda \cdot H (\theta_I, V) = 0
  \quad \mbox{where} \quad 
  H (\theta_I, V) = \sum_{k \geqslant 2} \frac{k}{k^2 - 1} \operatorname{Im} \left(
   \frac{\hat{f}_I (k)}{1 - 2 i \pi kV} \right) .
   \label{eq: H}
\end{equation}
Note that $f_I$ and hence $H$ depends on $\theta_I$, see
equation~(\ref{eq:fI}).

When $\theta_I = 0$, $f_I (\omega) = f (\omega)$ is an even function of
$\omega$, so that $\hat{f}_0 (k)$ is real, hence $H (0, 0) = 0$. It is also
clear from the form of $H (\theta_I, V)$ that $\dfrac{\partial H}{\partial
\theta_I}$ and $\dfrac{\partial H}{\partial V}$ are generally non-zero. By the
implicit function theorem, we can solve~(\ref{eq: H}) for $V = V (\theta_I)$,
at least for $\theta_I$ small enough. We do so numerically; the result is
shown in Figure~\ref{fig:rolling}(c) as the dashed line, and agrees well with
the non-linear simulations. In Figure~\ref{fig:rolling}(d), the distribution
of natural curvatures predicted by the linear theory is compared to the
non-linear numerical simulations, and a good agreement is obtained as 
well.

Remarkably, the intensity of illumination $\Lambda$ factors out in
equation~(\ref{eq: H}) selecting the rolling velocity, so that $V$ depends on
$\theta_I$ but not on $\Lambda$ in this linear theory: this explains why the
rolling velocity is largely independent of $\Lambda$ in the non-linear
simulations.

\section{Waves in doubly clamped beams}\label{sec: DoublyClampedFlapping}

The second example we study is motivated by the experiments of
Gelebart \emph{et al.}~{\cite{Gelebart2017}}.  These experiments were
done on a twisted nematic strip -- where the nematic directors are
aligned along the length of strip on one surface called the
\emph{planar face} and normal to the beam face on the other surface
\emph{homeotropic face}. The goal is to induce contraction on one face and expansion on the other in order to maximize the magnitude of the
photo-bending coupling $| \lambda |$.  Exposing the planar face to light
makes $\lambda < 0$ while exposing the homeotropic face to light makes
$\lambda > 0$. In view of the analysis done in Section~\ref{sec:
PhotoElasticity}, $\Lambda \propto \alpha \propto -\lambda$, so
illuminating the planar (respectively, homeotropic) face corresponds
to $\Lambda > 0$ (resp.~$\Lambda < 0$) in our model.  Illumination, either due to
the direct effect or due to temperature rise or both, reduces the
nematic order causing a contraction by $(r/r_0)^{2/3}$ when
illuminated on the planar face and an extension by $(r_0/r)^{1/3}$
when illuminated on the homeotropic face where $r$ (respectively
$r_0$) is the anisotropy parameter in the illuminated (respectively
ambient) state.  Since $r < r_0$, for fixed unscaled illumination intensity $I_0$, we expect the resulting photo-strain
and spontaneous curvature coefficients $0 \le \Lambda_p \approx -2
\Lambda_h$, where $\Lambda_p$ is the coefficient when illuminated on the planar side and $\Lambda_h$ when illuminated on the homeotropic side. This distinction between $\Lambda_p$ and $\Lambda_h$ is caused by the small penetration depth only activating the \emph{trans} to \emph{cis} isomerization on the illuminated side; therefore, it is only the nematic orientation on the illuminated surface that matters. We study the results of our model first, and compare to the experimental
observations next.

\begin{figure}
  \begin{center}
    \includegraphics[width=6.5in]{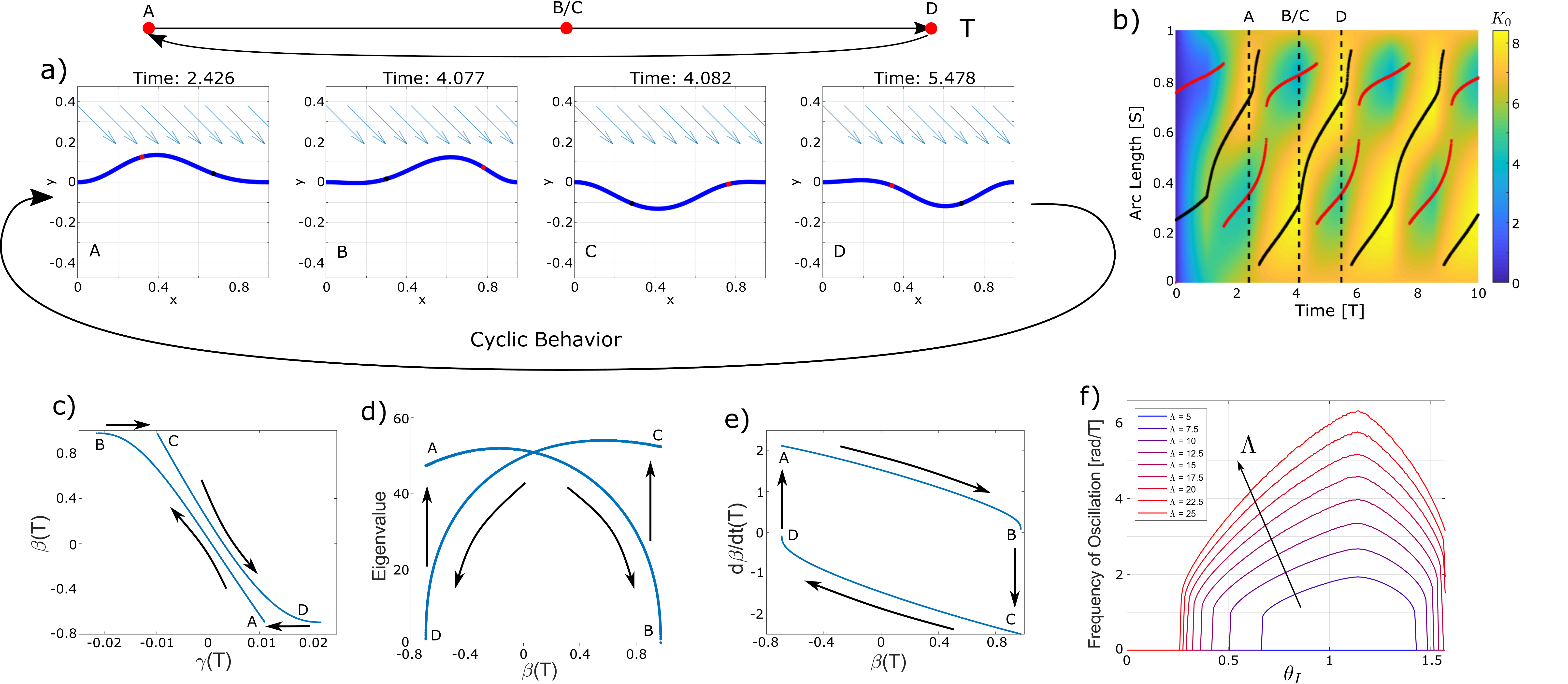}
  \end{center}
  \caption{Waves in a strip for $\Lambda > 0$. (a)~Snapshots of an initially flat
  strip clamped in a buckled state ($l_f = 0.95$) and subjected to
  illumination with $\Lambda = 10, \theta_I = \pi / 4$.  After an
  initial transient, it goes into a periodic motion.  (b)~Evolution of
  the light-induced spontaneous curvature $K_{0}$ of the strip.  The
  peaks are marked with a black curve and the troughs are marked with
  a red curve.  Note that the evolution becomes periodic but is quite
  complex with an alternation of slow (quasi-static) and fast
  (dynamic) motions.  (c)~Shape vs.~spontaneous curvature descriptors
  as defined in Equation~(\ref{eq:descriptors}).  (d)~Incremental
  stiffness (lowest eigenvalue of the stiffness matrix)
  vs.~spontaneous curvature descriptor. (e)~Phase plot revealing the oscillation cycles after an
  initial transient.  (f)~Frequency of flapping as a function of
  illumination angle for various illumination angles.
  \label{fig:flapping}}
\end{figure}

We first consider the case $\Lambda > 0$.  We take a strip that is
flat in the absence of any light or stress, so that $K_r = 0$.  We use
the same scaled quantities as earlier, and the scaled length of the
strip is 1.  We clamp the two ends at a distance $l_f < 1$ from each
other, corresponding to boundary conditions
\begin{equation}
  \label{eq:dcbc} \theta (0, T) = \theta (1, T) = 0, \quad \int_0^1 \sin
  \theta (S, T) \mathrm{d} S = 0 \hspace{0.17em}, \quad \int_0^1 \cos \theta (S,
  T) \mathrm{d} S = l_f .
\end{equation}
Since $l_f < 1$, the beam buckles and there are two equivalent
fundamental buckled modes -- buckled up and down.  We choose one of the
two states, say the buckled up state for definiteness, although the
results are independent of this choice.  We illuminate the strip with
a light source that is spatially uniform and at an angle ($\theta_I
\neq 0$) as shown in Figure~\ref{fig:flapping}(a).  We solve the
equations~(\ref{eq:equil}--\ref{eq:evol}) subject to the boundary
conditions~(\ref{eq:dcbc}).

Figure~\ref{fig:flapping}(a-e) show a typical simulation result.
After an initial transient, we find that the beam goes into a periodic
motion alternating between the up and down buckled shapes, see Figure
\ref{fig:flapping}(a).  At the start of the cycle, we have an up-bump
at the left side of the strip (state A).  Illumination moves it to the
right initially rapidly but slowing down and becoming very slow as it
reaches the right end (B).  It then suddenly changes shape and a down
bump appears on the left(C).  Subsequently, the down-bump moves to the
right initially rapidly but slowing down and becoming very slow as it
reaches the right end (D).  It then changes shape suddenly to the
starting point of the up-bump on the left, and the cycle repeats.

The evolution of the light-induced spontaneous curvature as a function of time and 
position is shown in Figure~\ref{fig:flapping}(b).  After an initial transient,
we see that the spontaneous curvature reaches a steady periodic cycle.
We also see that the spontaneous curvature does not
change during the sudden change from states B to C (and from D to A). This is
emphasized in Figure \ref{fig:flapping}(c) which plots one particular Fourier
component $\gamma (T)$ of the deflection, against one particular Fourier
component $\beta (T)$ of the natural curvature,
\begin{equation} \label{eq:descriptors}
  \gamma (T) = \int_0^1 \sin (2 \pi S) Y (S, T) \mathrm{d} S \hspace{0.17em},
  \quad \beta (T) = \int_0^1 \sin (2 \pi S) K_0 (S, T) \mathrm{d} S.
\end{equation}
We call these quantities the descriptors of the deformation and curvature, respectively. This suggests that the sudden changes from state B to C, and from D to A,
correspond to a snap-through bifurcation, from one equilibrium solution of the
elastica to another one. For some fixed time $T$ and spontaneous curvature
distribution $K_0 (S, T)$, the equilibrium equation~(\ref{eq:equil}) may have
multiple solutions (equivalently, $\mathcal{E}$ has multiple stationary
points). Stable solutions are those for which the second variation is positive
definite. With the aim to confirm the snap-through scenario, we study the
lowest eigenvalue associated with the second variation $\delta^2 \mathcal{E}$
of the energy. It is plotted from the numerical solution, as a function of
$\beta$ in Figure~\ref{fig:flapping}(d). We see that this eigenvalue is
positive at the start of the cycle at A (the solution with the up-bump) but
decreases as we go from A to B. The jump at B occurs when the eigenvalue
is becoming negative and the solution loses stability. It arrives on an other
solution C having a down-bump, which appears to be elastically stable,
{\emph{i.e.}},~has a positive lowest eigenvalue. Again, the lowest eigenvalue
begins to decrease as we go from C to D and passes through zero at D.

This reveals the mechanism of the cyclic motion.  At any time, there
are two possible solutions, one with an up-bump and one with a
down-bump.  If the solution with the up-bump has the bump on right,
the solution with the down-bump has the bump on the left and
vice-versa.  The evolution of light-induced spontaneous curvature
always forces the bump to the right, {\emph{i.e.}}, away from the
light source.  At some point it loses stability and has to snap to the
other solution.  The periodic cycles are represented in the phase
space $(\beta,\dot{\beta})$ in Figure \ref{fig:flapping}(e).
Immediately after a snap-through, the evolution speed $|\dot{\beta}|$
is high.  As the instability is approached, the magnitude of
$|\dot{\beta}|$ decreases until nearly zero.  This coincides with the
snap through and once the system snaps to the new configuration,
$|\dot{\beta}|$ jumps to a large value again, and the other half of
the cycle proceeds similarly.

We repeat this calculation for various illumination angles and
illumination intensities, and the results are summarized in
Figure~\ref{fig:flapping}(f).  At any given intensity, there is a
window of illumination angles at which periodic flapping solutions are
observed.  Outside this window, a stationary solution is reached,
which can be the up-bump or the down-bump depending on the initial
conditions.  This window becomes wider when the light intensity is
increased.  Further, at any given orientation, we see that the
frequency of the limit cycle increases with intensity.

\begin{figure}
  \begin{center}
    \includegraphics[width=6.5in]{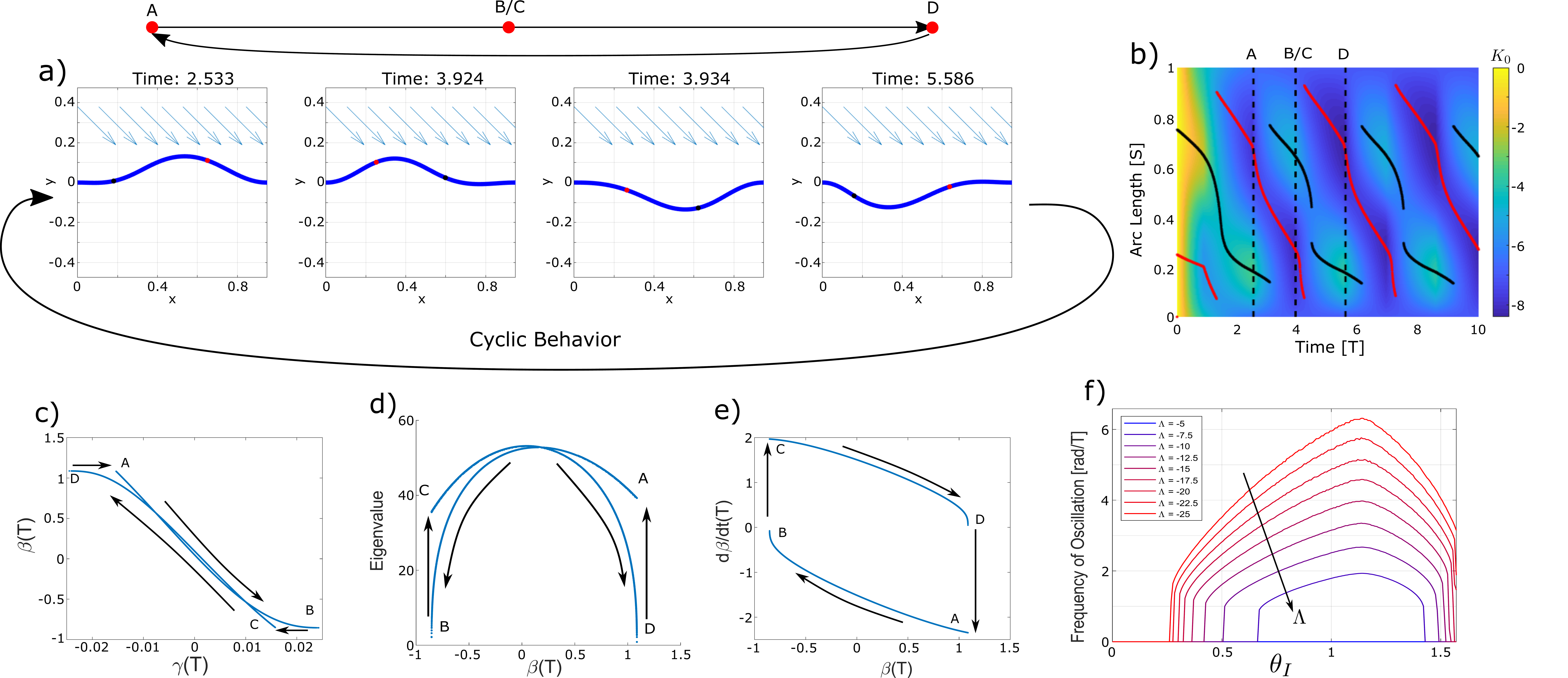}
  \end{center}
  \caption{Waves in a strip for $\Lambda < 0$. Same as in Figure~\ref{fig:flapping} except with $\Lambda = -10$.
  \label{fig:flappingNegative}}
\end{figure}

We now turn to the case when $\Lambda < 0$.  As can be seen in
Figure~\ref{fig:flappingNegative}(a), the system again alternates
between up and down buckled states.  In this case, however, the bulge
propagates from right to left (opposite from the case where $\Lambda >
0$).  It can be seen in (c)--(e) that the descriptors give different
paths through the phase space when compared to the case where $\Lambda
> 0$.  This shows that flipping the sign of $\Lambda$ does not simply
amount to reverse the arrow of time.  Interestingly, even though the
deformation mode differs, the flapping frequency (f) does not change
significantly between the positive and negative cases.

We now compare the experimental observations of Gelebart \emph{et al.}~{\cite{Gelebart2017}}. When illuminated on the homeotropic phase, $\Lambda = \Lambda_h <0$:
after an initial transient, the strip begins a periodic motion with
the wave moving from right to left predicted in Figure
\ref{fig:flappingNegative}.  When illuminated on the planar face where
$\Lambda = \Lambda_p > 0$, the wave moves from left to right as
predicted in Figure \ref{fig:flapping}. They also observed that the frequency of oscillation when illuminating the homeotropic face is lower as compared to the planar face, holding all other parameters fixed. Again, this is consistent with the
predictions in Figures \ref{fig:flappingNegative}(f) and
\ref{fig:flapping}(f) since $|\Lambda_p| > |\Lambda_h|$ for fixed $I_0$.  Further, this wave-like
motion is observed only in a finite range of illumination angles and, for fixed illumination intensity, the range when illuminating the planar side is larger than that of the homeotropic side as predicted because $|\Lambda_p| > |\Lambda_h|$. All these results are in good agreement with the experimental observations.

\paragraph{Acknowledgement} This work started while BA visited Caltech as a
Moore Distinguished Scholar in 2017-18.  We are pleased to acknowledge useful
discussions with Alexa Kuenstler and Ryan Hayward. KK and KB gratefully acknowledge the
support of the US Office of Naval Research through the MURI grant ONR
N00014-18-1-2624. KK also acknowledges the support of the National Science
Foundation Graduate Research Fellowship under Grant No. DGE-1745301.

\newpage

\bibliography{citations}

\newpage 

\begin{center}
{\Huge Supplementary Material}
\end{center}
\vspace{0.2in}

\appendix\section{Computational Model}\label{sec: CompMod}
\setcounter{footnote}{0}

The numerical method is motivated by the discrete elastic rod model\footnote{Mikl\'os Bergou, 
Max Wardetzky, Stephen Robinson, Basile Audoly, and Eitan Grinspun. 
Discrete elastic rods.  {\it ACM Transactions on Graphics}, 27(3):63:1 -- 63:12, August 2008.}.
We partition the beam into $N-1$ segments ${\mathcal S}^i = (S_i, S_{i+1}), \ i = 1, \dots, N-1$ 
which are all equal in 
arc-length by introducing $N$ nodes: the $i$th node is at arc-length $S_i = (i - 1) L / (N -
1)$.  We introduce the angle $\theta^i, \ i = 1, \dots, N-1$ to be the angle that the 
segment ${\mathcal S}^i$ makes to the horizontal as our main kinematic variable.
We can then obtain the current position of the $n$th node by exploiting 
the inextensibility condition as follows:
$$
 \mathbf{x}_n =  \mathbf{x}_1 + 
 \sum_{i = 2}^n (S_i - S_{i - 1})  \left( \cos \theta^{i - 1}
   \mathbf{e}_1 + \sin \theta^{i - 1} \mathbf{e}_2 \right).
 $$

The curvature is carried at the nodes and defined as $\kappa_i = \theta^i - \theta^{i-1}$
so that the total bending energy of the beam (discrete equivalent to (\ref{eq:be})) is given by
\begin{equation}
  \label{eq: BendingEnergy} E_B [\theta] = \sum_{i = 2}^{N - 1} \frac{1}{2}
  J_i  (\theta^i - \theta^{i - 1} - \kappa_i^0)^2
\end{equation}
where  $J_i$ is a
bending modulus and $\kappa_i^0$ is the discrete natural curvature at the
$i$th node.

We obtain the equilibrium equation (discrete equivalent to (\ref{eq:equil})) by taking the 
variation of $E_B$ with respect to $\theta^j$:
\begin{equation} \label{eq:elasticEquilibrium}
{\partial E_B \over \partial \theta^j} = 
J_j  (\theta^j - \theta^{j - 1}   - \kappa_j^0) - J_{j + 1}  (\theta^{j + 1} - \theta^j - \kappa_{j + 1}^0) = 0.
\end{equation}
Given the spontaneous curvatures $\{\kappa_j^0\}$, we solve these equations for 
$\{\theta^j\}$ subject to appropriate boundary conditions.
In order to improve the stability and convergence, it is convenient to have the Hessian,
$$
{\partial^2 E_B \over \partial \theta^j \partial \theta^k} = - J_j
  \delta_k^j + (J_j + J_{j + 1}) \delta_k^j - J_{j + 1} \delta_k^{j + 1} .
$$

It remains to specify the spontaneous curvature.  This evolves according to
(\ref{eq:evol}) whose discrete version is given by the set of ordinary differential equations:
\begin{equation}
  \dfrac{\mathrm{d} \kappa_i^0}{\mathrm{d} t} + \kappa_i^0 = \Lambda f (\theta_i -
  \theta_I), \label{eq:discreteIllumationEq}
\end{equation}
where $f (\theta_i - \theta_I)$ is as defined in Equation \ref{eq:evol} and $\theta_i = \text(ave)(\theta^i,\theta^{i-1}) $ is defined as the angle of the tangent of the $i$th node. 

Equation~(\ref{eq:discreteIllumationEq}) is discretized in time using an explicit Newton
time stepping algorithm. Time dependent solutions are obtained by alternating the elastic relaxation in
equation~(\ref{eq:elasticEquilibrium}) and evolving of natural curvatures
$\kappa_i^0$ based on equation~(\ref{eq:discreteIllumationEq}) over a time
step.

\subsection{Elastic Ring}\label{sec: ElasticRingComp}

In Section~\ref{sec:rolling}, we analyzed rolling rings. They can be simulated
by adapting the general numerical procedure outlined above as follows. The
closure of the ring is imposed by the following constraints:
\[ \theta^1 = 0 \qquad \mathbf{x}_1 = \mathbf{x}_{N - 1} \qquad \mathbf{x}_2 =
   \mathbf{x}_N \]
The first of these can be implemented explicitly by freezing that degree of freedom and represents that the point
of contact is tangent to the surface.

The last two enforce the closure constraint. The system is initialized by
assuming a constant curvature which makes the last two nodes coincident with
the first two. Then the system is relaxed by minimizing the energy while
imposing the constraints. In order to stabilize the point of contact when the
system is circular, a small amount of gravity is initially added and removed
once the natural curvature deviates from its initial state. 

The algorithm for calculating the translation and rotation of the system is as follows. Initially, the point of contact is defined to be the first and second nodes (second to last and last due to constraints). Then, given a natural curvature, $\kappa_j^0$, the energy is minimized to find the new configuration. The natural curvature is then updated using the explicit forward Euler scheme according to (\ref{eq:discreteIllumationEq}). Then, using a small window near the first and second nodes (which wraps around to nodes on the far end of the beam), the closest node to the calculated center of mass is found. Then, the nodes on either side of that node are tested to find the closest to the center of mass. This then forms an ordered pair of nodes $(\mathbf{x}_i, \mathbf{x}_{i+1})$ which defines the segment closest to the center of mass. Then, by shifting the minimized curvature $\theta^i \to \theta^1$, $\theta^{i+1} \to \theta^2$, etc in a cyclic manner (so the quantities at end points get wrapped around the beam). Similar transformations are done to the natural curvature ($\kappa_i^0 \to \kappa_{N-1}^0$, $\kappa_{i+1}^0 \to \kappa_2^0$). Note that these transformations are done in such a way that the ordering of the nodes is preserved and wrapped. At this point, the updated points of contact are now the $1$st and $2$nd nodes and the algorithm can be repeated to integrate the system in time. This solves for the rotation of the system while the translation can be found by using the rolling contact condition. Using the convention before, we had set $\mathbf{x}_{1} = \mathbf{0}$. We can set this to be the relative position where the true position of node $i$ is defined as $\tilde{\mathbf{x}}_i = \mathbf{x}_{S_c} + \mathbf{x}_i$ where $\mathbf{x}_{S_c}$ is the position of the point of contact. $\mathbf{x}_{S_c}$ is found using the rolling condition. Let $\mathbf{x}_{S_c}^k$ be the position of the point of contact at time step $k$ and $i$ be the shift necessary to establish that the point of contact is vertically aligned with the center of mass. Then, 
\[
\mathbf{x}_{S_c}^{k+1} = 
\begin{cases} \mathbf{x}_{S_c}^k + (S_i - S_1)\mathbf{E}_1 & \text{if } i \in [1,Ns] \\ 
 \mathbf{x}_{S_c}^k + (S_i - S_{N-1})\mathbf{E}_1 & \text{if } i \in [N-Ns,N-1] \, ,
 \end{cases}
\] 
where $N_s$ is a small window (usually set to $N/20$). If $i$ is not in the range of values defined above, then the time discretization is made finer in order to ensure that the rotations induced in each time step correlate with a small translation. The results for various angles of incidence of light and intensities are given in Movie S1 in the supplementary material.
The "velocity" of the system is then found by finding the distance the point of contact travels over a small time window. Steady state velocities are found by iterating the time stepping procedure until the velocity reaches a steady value.

\subsection{Doubly Clamped Beam}\label{sec: DoublyClampedComputation}

The doubly clamped system can be solved by setting up the following
constraints
\[ \theta^1 = 0 \qquad \theta^{N - 1} = 0 \qquad \mathbf{x}_N = l_f 
   \mathbf{e}_1 \hspace{0.17em} \]
where $l_f < L$ is the distance between the two endpoints. As before, the
first of these two constraints can be implemented explicitly by freezing those degrees of freedom and requires no
special treatment, while the latter two constraints need to be implemented in
the optimization engine. The initial solution is obtained numerically by
decreasing $l_f$ from 1 to its actual value in small steps. The system is integrated in time by alternating between relaxing the elastic energy and updating the natural curvature using an explicit Newton time stepping method. The results for various angles of incidence of light and intensities are given in Movie S2 in the supplementary material.

\section{Equilibrium and Stability Analysis}\label{sec: Stability}

Investigation of the snapping instabilities from Section~\ref{sec:
DoublyClampedFlapping} requires obtaining the second
variation of the energy ${\mathcal E} (\theta)$ in the presence of $m$ constraints $c_i
(\theta) = 0$, $i = 1, 2, ..., m$, where $\theta \in \mathbb{R}^n$ is the set
of degrees of freedom. Denote the feasible set $\mathcal{C}= \{\theta \in
\mathbb{R}^n  \text{s.t. } c_i (\theta) = 0\}$. We are interested in solutions
$\bar{\theta} \in \mathcal{C} \subset \mathbb{R}^n$ such that
\[ {\mathcal E} \left( \bar{\theta} + \varepsilon u + \frac{1}{2} \varepsilon^2 w \right)
   \geq {\mathcal E} (\bar{\theta}) \hspace{0.17em}, \quad \forall u, w \in \mathbb{R}^n
\]
satisfying $\bar{\theta} + \varepsilon u + \frac{1}{2} \varepsilon^2 w \in
\mathcal{C}$, with $\varepsilon \rightarrow 0$. Expanding each of these out to
first order and simplifying gives,
\[ \nabla {\mathcal E} (\bar{\theta}) \cdot u = 0 \hspace{0.17em}, \]
\[ \nabla c_i (\bar{\theta}) \cdot u = 0 \hspace{0.17em} . \]
where $\nabla$ denotes the gradient operator relative to the degrees of
freedom of the function ($(\nabla E)_i = \tfrac{\partial E}{\partial
\theta_i}$). This gives the equilibrium condition,
\[ \nabla {\mathcal E} (\bar{\theta}) + \sum_{i = 1}^m \lambda_i \nabla c_i
   (\bar{\theta}) = 0, \]
where the parameters $\lambda_i$ are Lagrange multipliers.

For stability, we require that any perturbation which satisfies the
constraints will increase the energy. To do this, we expand our system to
second order in $\varepsilon$ and simplify:
\[ u \cdot \nabla^2 {\mathcal E}(\bar{\theta}) u + \nabla {\mathcal E} (\bar{\theta}) \cdot w \geq
   0 \hspace{0.17em}, \]
\[ u \cdot \nabla^2 c_i (\bar{\theta}) u + \nabla c_i (\bar{\theta}) \cdot w =
   0 \hspace{0.17em}, \]
where $\nabla^2$ is the Hessian operator which returns the symmetric matrix of
second derivatives. Using the equilibrium condition, we have
\[ \nabla {\mathcal E} (\bar{\theta}) \cdot w = - \sum_{i = 1}^m \lambda_i \nabla c_i
   (\bar{\theta}) \cdot w = u \cdot \sum_{i = 1}^m \lambda_i \nabla^2 c_i
   (\bar{\theta}) u \hspace{0.17em} . \]
Plugging this into the above inequality, we have the stability condition that
\[ u \cdot \left( \nabla^2 {\mathcal E} (\bar{\theta}) + \sum_{i = 1}^m \lambda_i
   \nabla^2 c_i (\bar{\theta}) \right) u \geq 0 \hspace{0.17em}, \]
for all $u$ such that
\[ \nabla c_i (\bar{\theta}) \cdot u = 0 \hspace{0.17em} . \]
To determine whether a configuration satisfies this condition, we want to
project $\mathbb{R}^n$ onto the space tangent to the constraints. This is done
by a Gram-Schmidt process where
\[ v_1 = \frac{\nabla c_1 (\bar{\theta})}{\| \nabla c_1 (\bar{\theta})\|}
   \hspace{0.17em}, \]
\[ v_k = \frac{\nabla c_k (\bar{\theta}) - \sum_{i = 1}^{k - 1} (\nabla c_k
   (\bar{\theta}) \cdot v_i) v_i}{\| \nabla c_k (\bar{\theta}) - \sum_{i =
   1}^{k - 1} (\nabla c_k (\bar{\theta}) \cdot v_i) v_i \|} \hspace{0.17em},
\]
\[ P = I - \sum_{i = 1}^m v_i \otimes v_i \hspace{0.17em} . \]
The stability analysis then boils down to calculating the eigenvalues of $P
\left( \nabla^2 {\mathcal E} (\bar{\theta}) + \sum_{i = 1}^m \lambda_i \nabla^2 c_i
(\bar{\theta}) \right) P$. Due to the projection, there will be $m$ zero
eigenvalues and stability is implied when all other eigenvalues are greater
than zero. This analysis determines if there exists feasible paths which
locally lowers the energy; therefore, the existence of a non-positive
eigenvalue implies a loss of stability of the configuration.

\end{document}